\documentclass[%
 reprint,
superscriptaddress,
 amsmath,amssymb, 
prb,
]{revtex4-1}

\usepackage{graphicx}% Include figure files
\usepackage{dcolumn}% Align table columns on decimal point
\usepackage{bm}% bold math
\usepackage{graphicx}
\usepackage{bm}
\usepackage{amssymb}
\usepackage{dcolumn}
\usepackage{subfigure}
\usepackage{threeparttable}
\usepackage{multirow}
\usepackage[usenames,dvipsnames]{color}
\usepackage{wasysym}
\usepackage{txfonts}
\usepackage{mathtools}

\begin{document}

\newcommand{\vn}[1]{{\bf{#1}}}
\newcommand{\vht}[1]{{\boldsymbol{#1}}}
\newcommand{\matn}[1]{{\bf{#1}}}
\newcommand{\matnht}[1]{{\boldsymbol{#1}}}
\newcommand{\bege}{\begin{equation}}
\newcommand{\ee}{\end{equation}}

%\preprint{APS/123-QED}

\title[Zhang \textit{et al}.]{All electrical manipulation of magnetization dynamics in a ferromagnet by antiferromagnets with anisotropic spin Hall effects}% Force line breaks with \\
%\thanks{Footnote to title of article.}

\author{Wei Zhang}
\email{zwei@anl.gov.}
\affiliation{Materials Science Division, Argonne National Laboratory, Argonne IL 60439, USA}

\author{Matthias ~B.~Jungfleisch}
\affiliation{Materials Science Division, Argonne National Laboratory, Argonne IL 60439, USA}

\author{Frank Freimuth}
\affiliation{Peter Gr\"unberg Institut and Institute for Advanced Simulation, Forschungszentrum J\"ulich and JARA, D-52425, J\"ulich, Germany}

\author{Wanjun~Jiang}
\affiliation{Materials Science Division, Argonne National Laboratory, Argonne IL 60439, USA}

\author{Joseph Sklenar}
\affiliation{Department of Physics and Astronomy, Northwestern University, Evanston IL 60208, USA}

\author{John E. Pearson}
\affiliation{Materials Science Division, Argonne National Laboratory, Argonne IL 60439, USA}

\author{John B. Ketterson}
\affiliation{Department of Physics and Astronomy, Northwestern University, Evanston IL 60208, USA}

\author{Yuriy Mokrousov}
\affiliation{Peter Gr\"unberg Institut and Institute for Advanced Simulation, Forschungszentrum J\"ulich and JARA, D-52425, J\"ulich, Germany}

\author{Axel Hoffmann}
\affiliation{Materials Science Division, Argonne National Laboratory, Argonne IL 60439, USA}

\date{\today}% It is always \today, today,

\begin{abstract}
We investigate spin-orbit torques of metallic CuAu-I-type antiferromagnets using spin-torque ferromagnetic resonance tuned by a dc-bias current. The observed spin torques predominantly arise from diffusive transport of spin current generated by the spin Hall effect. We find a growth-orientation dependence of the spin torques by studying epitaxial samples, which may be correlated to the anisotropy of the spin Hall effect. The observed anisotropy is consistent with first-principles calculations on the intrinsic spin Hall effect. Our work demonstrates large tunable spin-orbit effects in magnetically-ordered materials. 
\end{abstract}

\maketitle

\section{\label{sec:level1}Introduction}

Ferromagnet/antiferromagnet (AF) bilayers have been core components in modern magnetic storage devices such as spin-valve structures and magnetic tunnel junctions, in which the antiferromagnets provide pinning for a reference ferromagnetic layer due to an interfacial effect called 'exchange-bias' \cite{jmmm_nogues}.  Exotic magnetic properties from such unidirectional pinning effect have been extensively studied in the past decades. Recent work shows also promising spin-orbit effects in antiferromagnets  \cite{haney_prl,xu_prl,nunez_prb,shick_prb,hals_prl,chengr_prl,zelezny_prl,shindou_femn,chen_irmn3,smzhou_prb2015,freimuth_prl2010} as well as efficient spin transfer via antiferromagnetic spin waves \cite{takei_prb2015,wanghl_prb2015,nio_prl,nio_epl,moriyama_apl2015,qu_prb2015}, 
enabling a more active role of antiferromagnets in the manipulation of ferromagnets beyond just a pinning effect. One particular example is the electrical manipulation of ferromagnets using spin-orbit effects, such as the spin Hall effect (SHE) \cite{yakonov,sinova_rmp}. The efficiency of the spin Hall effect can be characterized by the spin Hall angle ($\theta_\mathrm{SH}$) \cite{hoffmann_ieee,wz_jap2015}, which is typically determined by the intrinsic spin-orbit coupling of the materials involved \cite{guo_prl,wanghl_prl}, and therefore cannot readily be varied by additional external parameters. Thus, much effort has focused on the extensive exploration of the right materials with large intrinsic spin Hall effect. Recently, it was found that the magnetic-proximity-induced magnetization states of heavy metals (Pt and Pd) also affects their intrinsic spin Hall effect \cite{wz_proximity}; therefore, magnetically ordered materials may offer additional opportunities to tune the intrinsic spin Hall effect via their atomic spin magnetic moments. CuAu-I-type antiferromagnetic alloys, such as PtMn, consisting of both heavy-metal elements (Pt) and atomic-level, staggered magnetization (Mn), may be promising candidates for efficient and tunable electrical manipulation of  ferromagnets \cite{wz_prl}. It should be also noted that in antiferromagnets with specific crystal symmetries, it is even possible to manipulate the antiferromagnetic spin configuration with electric currents via intrinsic spin-orbit torques \cite{wadley}. Last but not least, the complementary spin-orbit effect and exchange-bias effect from a single material may also enable new device functionalities. 

In this work, we use an electrical detection technique of the ferromagnetic resonance of Permalloy (Py = Ni$_\mathrm{80}$Fe$_\mathrm{20}$) driven by the in-plane ac-current from four CuAu-I-type antiferromagnets (AF = PtMn, IrMn, PdMn, and FeMn). \textcolor{black}{The experimental details are discussed in Section II.} \textcolor{black}{In Section III-A,} we show that antiferromagnets can serve as an efficient spin current source that can be used to manipulate the magnetization in ferromagnets, as illustrated in Fig.1(a). Apart from the fact that appreciable spin Hall effect originates from the large, atomic scale spin-orbit coupling of the heavy elements, the staggered magnetization of Mn may also play important role for their spin Hall effects as revealed by epitaxial  samples \textcolor{black}{(Section III-B), whose significance is further corroborated by first-principles calculations (Section III-C).} The efficient generation of spin current, together with other advantages of antiferromagnets including insensitivity to external field, lack of stray fields, faster spin-dynamics, and effective spin current transmission, will pave the way for future antiferromagnetic-based spin-orbitronics \cite{park_tamr,mendes_rapid,marti_nmat,fina_ncomm,wangyy_prl,bailey_apl,tsoi_prx}. 

\begin{figure}[b]
\includegraphics[width=1\columnwidth]{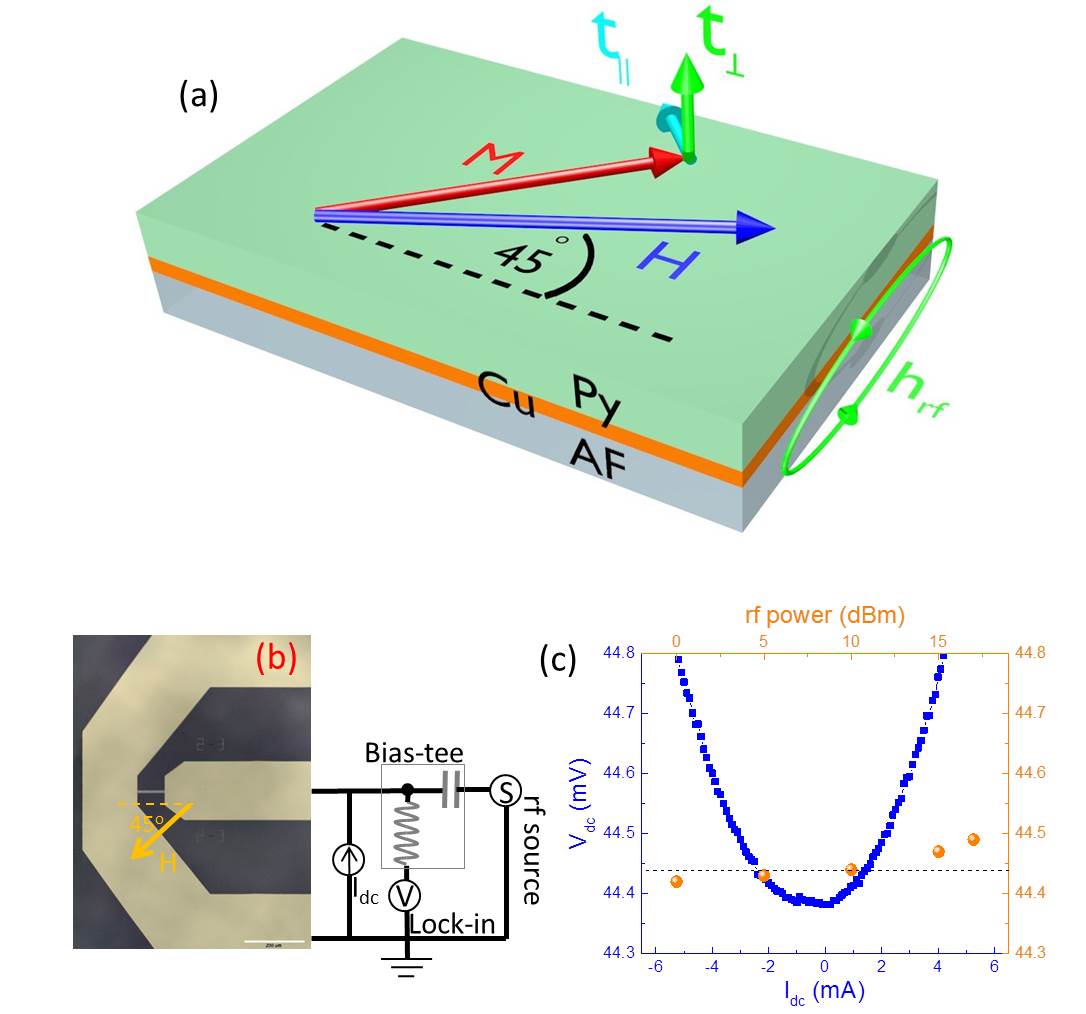}
\caption{\label{fig1} (Color online) (a) Schematic illustration of the spin Hall effect induced spin transfer torques ($\tau_{||}$ and $\tau_\perp$) of AF/Cu/Py multilayers. (b) Depiction of the circuit used for the spin-torque ferromagnetic resonance measurement and the sample contact geometry. (c) Comparison of resistance change due to heating caused by dc and rf currents.}
\end{figure}

\section{\label{sec:level1}Experiments}

All our samples, having the structures of AF($t_\mathrm{AF}$=10)/Cu($t_\mathrm{Cu}$=1)/Py($t_\mathrm{Py}$=5) or AF(10)/Py(5) [thicknesses in nm], were deposited on 1 cm $\times$ 1 cm MgO(001) substrates by magnetron sputtering at rates $<$ 1 $\AA$/s. The 10 nm thicknesses of the antiferromagnetic layers ensure their magnetically ordered states. Polycrystalline samples were grown at room temperature (RT) and epitaxial ones were grown at elevated temperatures. Cu and Py film stacks were subsequently grown \textit{in-situ} after cooling down the antiferromagnetic films to minimize interdiffusion and to ensure identical growth enviroment for Cu and Py. The multilayer film stacks were microstructured into microstrips, with varying lengths (25 - 90 $\mu$m) and widths (5 - 20 $\mu$m). Ti/Au ground-signal-ground electrodes were patterned using photolithography and lift-off [Fig. 1(b)].

\begin{figure}[t]
\includegraphics[width=1.05\columnwidth]{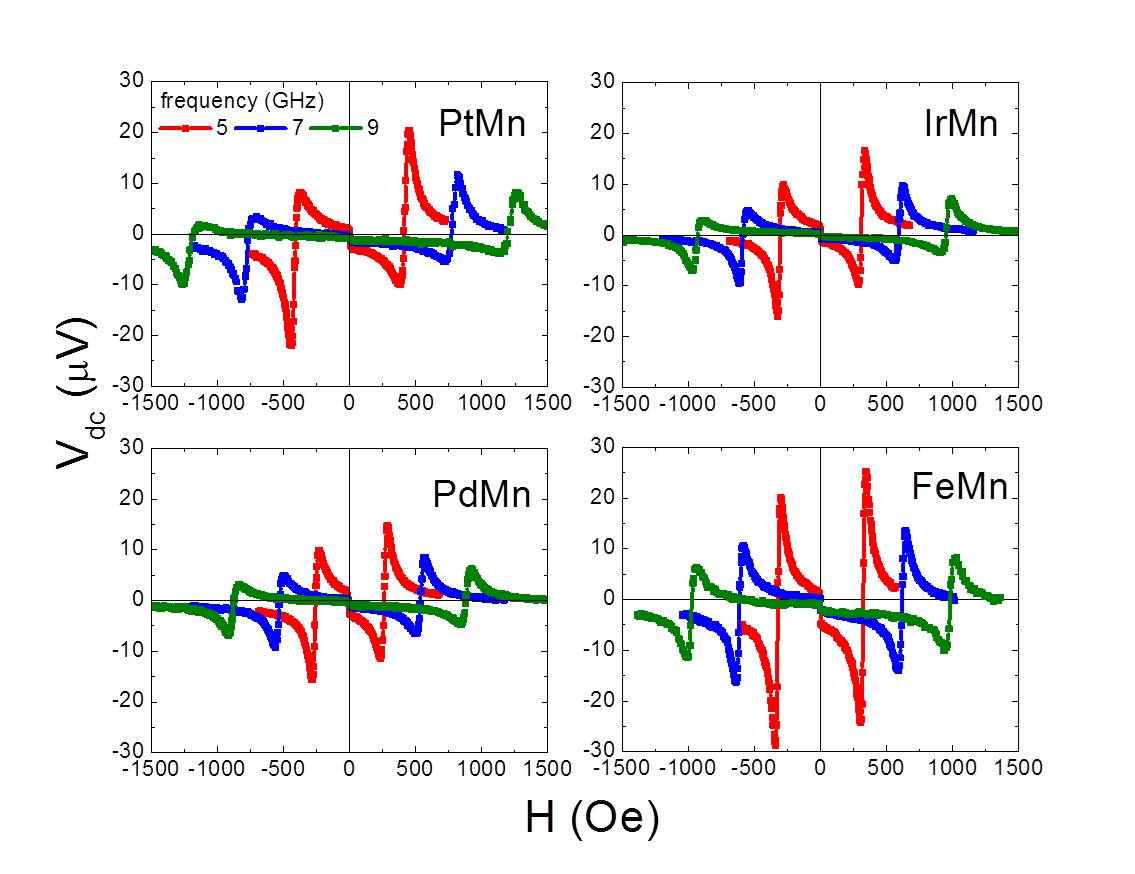}
\caption{\label{fig2} (Color online) Measured spin-torque ferromagnetic resonance signals from AF(10)/Cu(1)/Py(5) at room temperature. }
\end{figure}

\begin{figure*}[t]
\includegraphics[width=2.1\columnwidth]{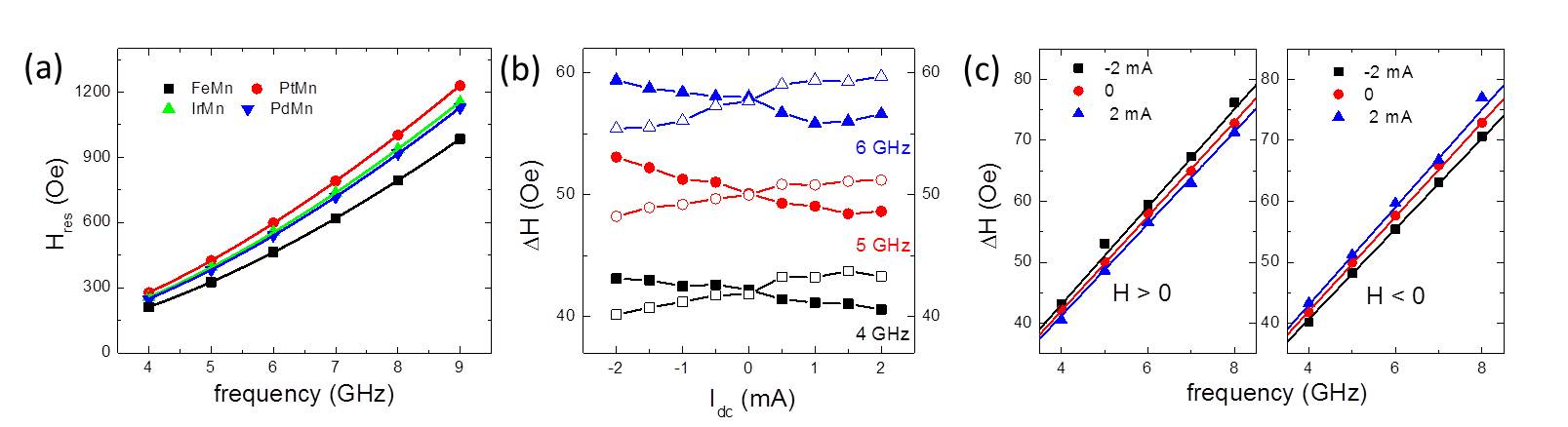}
\caption{\label{fig3} (Color online) (a) Resonance field versus frequency and the corresponding Kittel fit for different antiferromagnetic sample stacks. (b) and (c) Resonance linewidth ($\Delta H$) versus dc-bias current ($I_\mathrm{dc}$) at different frequency, \textit{f} = 4, 5, 6 GHz [dc current polarity: solid (+) empty (-)], and resonance linewidth versus frequency at different dc-bias current, $I_\mathrm{dc}$ = -2, 0, 2 mA, for PtMn(10)/Py(5).}
\end{figure*}

Spin-torque ferromagnetic resonance measurements \cite{liu_science} with a dc-current ($I_\mathrm{dc}$) tuning technique \cite{liu_prl2011} were performed for all samples. We apply microwave electrical currents at fixed frequency (4 -- 9 GHz) to the microstrips and sweep the magnetic field, \textit{H}, applied along $\phi$ = 45$^{\circ}$ with respect to the long axis of the device, as shown in Fig.1(b). The torques induced by the oscillating current drive magnetization precession of the Py, which is detected as a rectified dc voltage ($V_\mathrm{dc}$) due to anisotropic magnetoresistance. The applied rf power is between +10 and +15 dBm. To calibrate the rf current we make use of the change of resistance from Joule heating. We first measure the resistance change due to dc heating, and then calibrate the rf current ($I_\mathrm{rf}$) which is $\sqrt{2}$ times the dc current under the same amount of Joule heating [Fig. 1(c)]. The rf current differs from device to device in the range of 1 - 2.5 mA. The resistivities of the antiferromagnets grown on MgO are calibrated using independent four-point measurements, \textcolor{black}{yielding 164.5, 272.3, 220.0, and 161.5 $\mu \Omega$ cm, for PtMn, IrMn, PdMn, and FeMn, respectively.} The resistivity of Py grown on the antiferromagnets was determined to be $\sim$ 54.4 $\mu \Omega$ cm. This value can be slightly higher for Py grown on AF/Cu, confirming that the Py can have different resistivities depending on the seed layer. With the knowledge of both the total rf current and the individual resistivity we can estimate how much rf current flows through each layer in the multilayer stacks, which is important information for analyzing spin-torque ferromagnetic resonance measurements quantitatively. 

Figure 2 shows the measured spin-torque ferromagnetic resonance signals from AF(10)/Cu(1)Py(5) at room temperature. The magnitude of the voltages is significantly higher than that for the pure Py reference sample (below $\sim$ 4 $\mu$V). \textcolor{black}{In the Py reference samples the measured voltage can be attributed to the spin rectification \cite{Juretschke1960,Juretschke1963,gui_prl} or magnonic charge pumping of Py \cite{azevedo_prb}}. $V_\mathrm{dc}$ can be fitted  by a sum of symmetric, $F_\mathrm{sym}(H)$ and antisymmetric, $F_\mathrm{asym}(H)$ Lorentzian functions, $V_\mathrm{dc} = V_\mathrm{s} F_\mathrm{sym}(H) +  V_\mathrm{a} F_\mathrm{asym}(H)$, in which the symmetric component, $V_\mathrm{s}$, and antisymmetric component, $V_\mathrm{a}$ correspond to the in-plane ($\tau_\mathrm{||}$) anti-damping-like and out-of-plane ($\tau_\mathrm{\perp}$) field-like torques, respectively. 

\section{\label{sec:level1}Results and Discussions}

\subsection{\label{sec:level1}Spin-orbit torques from antiferromagnets}

The spin Hall angle can be quantified from two different methods by the lineshape analysis and also from analysis of the dc current dependence:

\subsubsection{\label{sec:level1}Ratio analysis}

The first method is from the ratio of the two voltages, $V_\mathrm{s}/V_\mathrm{a}$ \cite{liu_prl2011}: 

\begin{equation}
\label{ratio}
\theta_\mathrm{SH} = \frac{V_\mathrm{s}}{V_\mathrm{a}} \frac{e\mu_\mathrm{0}M_\mathrm{s}t_\mathrm{AF}t_\mathrm{Py}}{\hbar}[1+\frac{4\pi M_\mathrm{eff}}{H}]^\frac{1}{2},
\end{equation} 
where $\mu_\mathrm{0}$ is the permeability in vacuum, $M_\mathrm{s}$ is the saturation magnetization, and $M_\mathrm{eff}$ is the effective magnetization of Py. $M_\mathrm{s}$ and $M_\mathrm{eff}$ are indistinguishable for Py with strong in-plane anisotropy. A fit of the resonance field, $H_\mathrm{res}$ versus frequency to the Kittel equation gives the values of $M_\mathrm{eff}$ for different samples [Fig. 3(a)]. However, this method assumes that $V_\mathrm{a}$ is only attributed to out-of-plane torques due to the Oersted field.

\subsubsection{\label{sec:level1}Individual lineshape analysis}

Another method analyzes the individual voltage amplitude and gives the torque values via \cite{mellnik_nature}: 

\begin{equation}
\label{Vs}
V_\mathrm{s} = - \frac{I_\mathrm{rf}\gamma \mathrm{cos}\phi}{4} \frac{dR}{d\phi} \tau_\mathrm{||}\frac{1}{\Delta}F_\mathrm{sym}(H), 
\end{equation}
and  
\begin{equation}
\label{Va}
V_\mathrm{a} = - \frac{I_\mathrm{rf}\gamma \mathrm{cos}\phi}{4} \frac{dR}{d\phi} \tau_\mathrm{\perp} \frac{(1+\frac{\mu_\mathrm{0} M_\mathrm{eff}}{H})^\frac{1}{2}}{\Delta}F_\mathrm{asym}(H), 
\end{equation}
where $\gamma$ is the gyromagnetic ratio, ${dR}/{d\phi}$ is the angular dependent magnetoresistance at $\phi$ = 45$^{\circ}$, and $\Delta$ is the frequency linewidth of the signal. $\theta_\mathrm{SH}^*$ can be obtained via: $\theta_\mathrm{SH}^* = \frac{\tau_\mathrm{||}M_\mathrm{s}t_\mathrm{Py}}{\sigma E}$, where $\sigma$ is the electrical conductivity of the antiferromagnet and \textit{E} the electric field from $I_\mathrm{rf}$. 

To extract the intrinsic spin Hall angles free from any possible interfacial exchange effects, we focus primarily on the samples with an atomically thin dusting layer made from Cu \textcolor{black}{which induces minimal current shunting. The thin layer of Cu breaks most of the interface exchange bias effect; however, a small unidirectional anisotropy may remain depending on the strength of the exchange bias as well as other microsopic details at the interface \cite{geshev_prb}.} The spin Hall angles extracted from the two lineshape methods (\textit{1} and \textit{2}) and for the four different antiferromagnets are summarized in Table I. PtMn shows the largest spin Hall angle, followed by IrMn, PdMn, and FeMn. This trend is in good agreement with previous spin pumping experiments \cite{wz_prl}, confirming the reciprocity between the spin-torque (driven by spin Hall effect) and the spin pumping (detected with the inverse spin Hall effect), although the absolute $\theta_\mathrm{SH}$ values are slightly different using the two experimental techniques.

\subsubsection{\label{sec:level1}dc-tuned damping analysis}

Alternatively, tuning the linewidth via additional dc currents also gives an effective spin Hall angle, $\theta_\mathrm{DL}$ \cite{nan_prb}. As shown in Fig.3(b), the linewidth ($\Delta H$) is reduced for one current polarity and enhanced for the opposite polarity, indicating a modulation effect of the empirical damping parameter, $\alpha_\mathrm{eff} = |\gamma|/2\pi f(\Delta H - \Delta H_\mathrm{0})$, via the damping-like torque [Fig. 3(c)]. Therefore, $\theta_\mathrm{DL}$ is calculated from the $I_\mathrm{dc}$ dependence of $\alpha_\mathrm{eff}$: 

\begin{equation}
\label{DL}
|\theta_\mathrm{DL}| = \frac{2e}{\hbar} \frac{(H_\mathrm{res}+\frac{M_\mathrm{eff}}{2}) \mu_\mathrm{0} M_\mathrm{s} t_\mathrm{Py}}{\mathrm{sin} \phi}|\frac{\Delta \alpha_\mathrm{eff}}{\Delta J_\mathrm{dc}}|, 
\end{equation}
where $J_\mathrm{dc}$ is the dc charge-current density. 

\begin{table}[b]
\begin{tabular*}{3.2in}[t]{ccccc}
 \hline
 \ ~~\ & PtMn \ & IrMn \ & PdMn \ & FeMn \\
  \hline
\ $\theta_\mathrm{SH}$ \ & 0.081$\pm$0.005 \ & 0.053$\pm$0.004 \ & 0.049$\pm$0.003 \ & 0.028$\pm$0.005 \\
\ $\theta_\mathrm{SH}^*$ \ & 0.064$\pm$0.005 \ & 0.057$\pm$0.004 \ & 0.028$\pm$0.004 \ & 0.022$\pm$0.003 \\
  \hline
\end{tabular*}
\caption{\label{table} Spin Hall angles obtained from the ratio ($V_\mathrm{s}/V_\mathrm{a}$) analysis, $\theta_\mathrm{SH}$, and from the voltage amplitude ($V_\mathrm{s}$) analysis, $\theta_\mathrm{SH}^*$, respectively.} 
\end{table}

\begin{figure}[t]
\includegraphics[width=1.05\columnwidth]{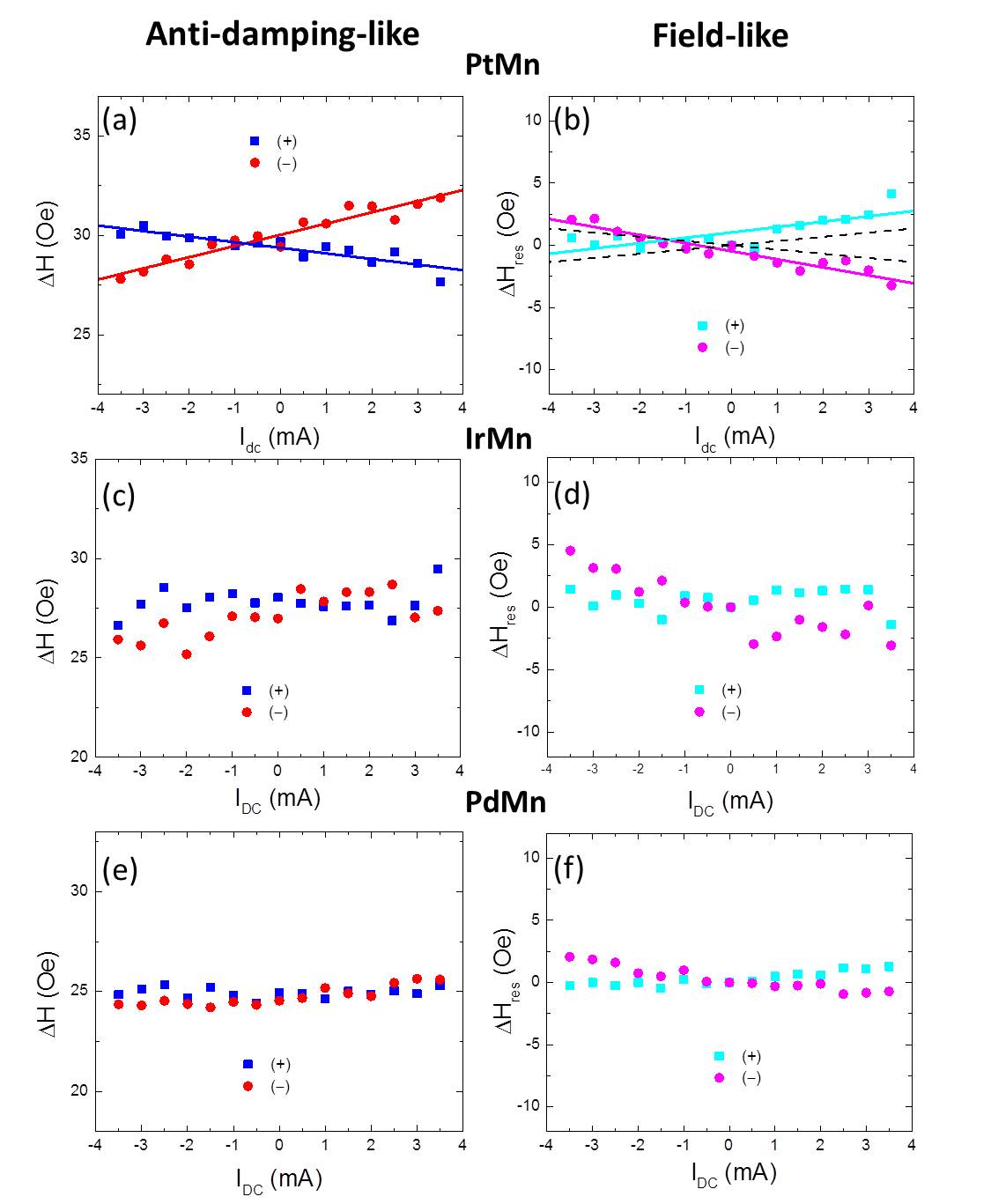}
\caption{\label{fig2} (Color online) (a,c,e) Resonance linewidth and (b,d,f) resonance field versus dc-bias current at 5 GHz for 5$\mu$m-wide PtMn/Cu/Py, IrMn/Cu/Py, and PdMn/Cu/Py devices. \textcolor{black}{Linear fitting for the dc-current effect is shown for PtMn. The calculated Oersted field contribution is indicated by the dashed line in (b). }}
\end{figure}

For the dc-current-induced linewidth modulations, however, we are only able to observe appreciable effect for the PtMn series of samples (Fig. 4). We obtain an anti-damping effective spin Hall angle for PtMn, $\theta_\mathrm{DL}$ = 0.079 $\pm$ 0.005, which is similar to the value obtained from the ratio analysis. On the other hand, the dc-induced shift in the resonance field gives us an estimate for the field-like torque. Such torque has the same polarity shift in $H_\mathrm{res}$ as the Oersted field, $\mu_\mathrm{0}H_\mathrm{Oe} = 0.048$ mT/mA based on the estimated charge-current densities in the antiferromagnetic and Cu layers: $H_\mathrm{Oe} = (J_\mathrm{dc, PtMn} t_\mathrm{PtMn} + J_\mathrm{dc, Cu} t_\mathrm{Cu})$/2. \textcolor{black}{As shown in Fig. 4(b)}, a fit for the $H_\mathrm{res}(I_\mathrm{dc})$ yields a total effective field, $\sqrt{2}\mu_\mathrm{0}H_\mathrm{res}$ = 0.061 \textcolor{black}{$\pm$ 0.012 mT/mA for the (+)-current and $0.092 \pm 0.008$ mT/mA for the (--)-current. We focus here on the linear effect of the dc-current modulation and neglect the asymmetrical behavior between (+) and (--). Such unidirectional asymmetry could be attributed to interfacial exchange-bias effect even through the 1 nm Cu layer \cite{geshev_prb}, which we do not aim to study in the present work. The calculated total effective fields} indicate the presence of an additional field-like torque per current, $\mu_\mathrm{0}H_\mathrm{FL} = \sqrt{2}\mu_\mathrm{0}H_\mathrm{res} - \mu_\mathrm{0}H_\mathrm{Oe}$, \textcolor{black}{in the range of} 0.013 \textcolor{black}{-- 0.044} mT/mA from PtMn. Therefore, a field-like effective spin Hall angle can be obtained by: 

\begin{equation}
\label{FL}
|\theta_\mathrm{FL}| = \frac{2 e \mu_\mathrm{0} M_\mathrm{s} t_\mathrm{Py}}{\hbar} |\frac{H_\mathrm{FL}}{J_\mathrm{dc, PtMn}}|, 
\end{equation}
yielding \textcolor{black}{at least} $\theta_\mathrm{FL}$ = 0.020 $\pm$ \textcolor{black}{0.004 for PtMn/Cu/Py using the lower end of the $H_\mathrm{FL}$ above}. The real and imaginary spin mixing conductance can then be calcuated by: 

\begin{equation}
\label{Re}
\mathrm{Re}[G_{\uparrow\downarrow}^\mathrm{eff}]=\frac{2e^2M_\mathrm{s}t_\mathrm{Py}}{\hbar^2 \gamma} (\alpha-\alpha_\mathrm{0})
\end{equation}
 and Im$[G_{\uparrow\downarrow}^\mathrm{eff}]$ = ($\theta_\mathrm{FL}/\theta_\mathrm{DL}$)Re$[G_{\uparrow\downarrow}^\mathrm{eff}]$. Using a pure Py(5) sample ($\alpha_\mathrm{0} = 0.01$), the calculations yield a Re$[G_{\uparrow\downarrow}^\mathrm{eff}]$ = (3.9$\pm$0.5) $\times$ 10$^{14}$ $\Omega^{-1}$m$^{-2}$ and \textcolor{black}{a minimum} Im$[G_{\uparrow\downarrow}^\mathrm{eff}]$ = (1.0$\pm$0.2) $\times$ 10$^{14}$ $\Omega^{-1}$m$^{-2}$. \textcolor{black}{The Im$[G_{\uparrow\downarrow}^\mathrm{eff}]$ is usually associated with the phase shift of the spin-orbit torques to the driving microwave, which can become quite prounounced in magnetically ordered materials \cite{smzhou_prb2015}.}

\textcolor{black}{Any experimentally-determined spin Hall angles, either using the lineshape or the damping analysis, are `effective' spin Hall angles bonded to the quality of the interface of the samples studied. Such interface properties depend on the materials, growths, crystallography, roughnesses and so on, which can vary largely for different samples. \cite{nan_prb,parkin_nphys,pai_apl,pai_prb,tokac_prl} In this regards, the `interface transparency' has been introduced to properly correct the interface effect which further allows the determination of an `internal' spin Hall angle, $\theta_\mathrm{SH}^\mathrm{int}$, of the materials, via: \cite{nan_prb,parkin_nphys,pai_prb}}

\begin{equation}
\label{internal}
\textcolor{black}{\theta_\mathrm{SH}^\mathrm{int}=\frac{\sigma / \lambda}{2 \mathrm{Re}[G_{\uparrow \downarrow}^\mathrm{eff}]} \theta_\mathrm{SH} \: (\mathrm{or} \: \theta_\mathrm{SH}^* \: \mathrm{or} \:  \theta_\mathrm{DL}),}
\end{equation}
\textcolor{black}{where $\sigma$ and $\lambda$ are the electrical conductivity and spin diffusion length of the spin Hall metal, respectively. The inverse of the prefactor, \textit{i.e.}, 2Re[$G_{\uparrow \downarrow}^\mathrm{eff}$]/($\sigma / \lambda$), is introduced as the spin current `transmissivity', \textit{T}. According to Eq. (7), \textit{T} is very sensitive to the spin diffusion length, which in itself is a material-dependent parameter that requires careful calibration for many spin Hall metals \cite{wz_jap2015,wz_apl,liu_review,niimi_prl,kimura_prl,isasa_prb,boone_jap2015}. Increasing the spin diffusion length linearly enhances the value of \textit{T} and decreases the value of $\theta_\mathrm{SH}^\mathrm{int}$.} Using the above Re$[G_{\uparrow\downarrow}^\mathrm{eff}]$ value and spin diffuson length of PtMn ($\lambda_\mathrm{PtMn}$ = 0.5 nm \cite{wz_prl}), the spin current transmissivity, \textit{T}= 2Re$[G_{\uparrow\downarrow}^\mathrm{eff}]$/($\sigma_\mathrm{PtMn}/\lambda_\mathrm{PtMn}$) and the internal spin Hall angle can be estimated. We obtain \textit{T} = 0.63 and an internal $\theta_\mathrm{SH}^\mathrm{int}$= 0.125 for PtMn, which exceeds the spin Hall angle of many reported paramagnetic metals \cite{hoffmann_ieee}.

\subsection{\label{sec:level1}Anisotropic spin-Hall effect in antiferromagnets}

\begin{figure}[b]
\includegraphics[width=1\columnwidth]{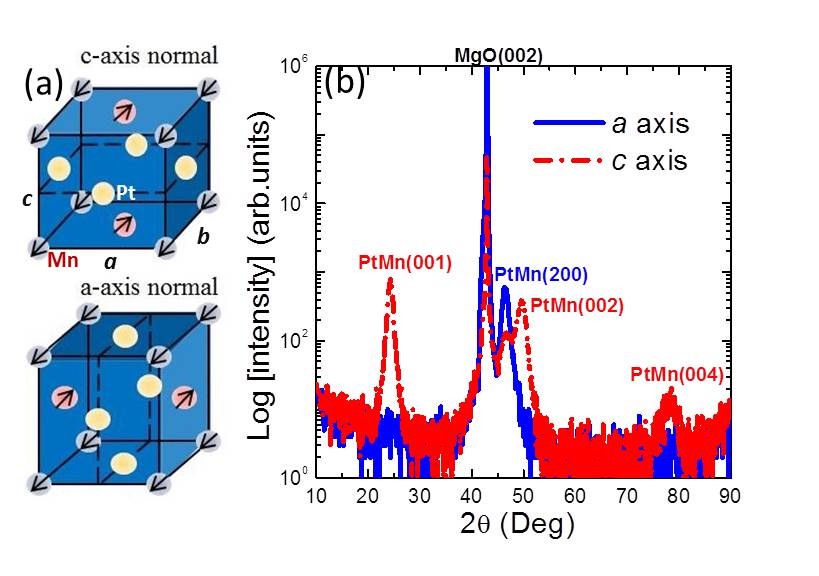}
\caption{\label{fig5} (Color online)  (a) A sketch illustrating the chemical structure of CuAu-I-type antiferromagnets grown along (001)[\textit{c}-axis normal] and (100)[\textit{a}-axis normal] directions, and (b) their corresponding x-ray diffraction patterns. The principal axes are inequivalent (\textit{c} $\neq$ \textit{a}) for PtMn, IrMn, and PdMn except for FeMn (\textit{c} = \textit{a}).}
\end{figure}

An important element of these antiferromagnetic alloys is the staggered magnetization ($M_\mathrm{stag}$) from Mn atoms that is strongly correlated to their crystal growths \cite{wz_AFs}. In the isotropic case the introduction of staggered magnetization (antiferromagnetic states) will break the symmetry and make it anisotropic. As a consequence, when the staggered magnetization is along a well-defined direction and not averaged, the anisotropy of the spin Hall effect in the CuAu-I-type antiferromagnets will arise both from the inequivalency of \textit{c/a} ratio (chemical structure) and the staggered magnetization (magnetic structure). 

We performed the same measurements and analysis on epitaxial samples for AFs with inequivalent \textit{c} and \textit{a} (lattice constants), \textit{i.e.}, excluding FeMn. The samples were grown at elevated temperatures following established recipes \cite{wz_AFs}. For example for PtMn, the \textit{c}-axis samples, following MgO(001)$||$PtMn(001), were grown at 550$^{\circ}$C; the \textit{a}-axis samples, following MgO(001)$||$PtMn(100), were grown at 120$^{\circ}$C and subsequently annealed at 250$^{\circ}$C for 1.5 hour. Their corresponding x-ray diffraction patterns are shown in Fig. 5. We obtain epitaxial structures for all materials except for \textit{a}-axis IrMn. The spin Hall angles estimated from the ratio analysis are summarized in Table II. 

\begin{table}[h!]
\begin{tabular*}{2.6in}[t]{cccc}
 \hline
 \ $\theta_\mathrm{SH}$\ & PtMn \ & IrMn \ & PdMn \\
  \hline
\ \textit{c}-axis \ & 0.052$\pm$0.002 \ & 0.050$\pm$0.005 \ & 0.032$\pm$0.006  \\
\ \textit{a}-axis \ & 0.086$\pm$0.002 \ & 0.023$\pm$0.005$^\dag$ \ & 0.039$\pm$0.005  \\
  \hline
\end{tabular*}
\caption{\label{table} Spin Hall angles, $\theta_\mathrm{SH}$, estimated from the ratio analysis ($V_\mathrm{s}/V_\mathrm{a}$) for epitaxial PtMn, IrMn, and PdMn. $^\dag$ The \textit{a}-axis IrMn is weakly textured and can be considered almost polycrystalline. } 
\end{table}

\subsection{\label{sec:level1}First-principles calculations}

To verify our assumption we have performed first-principles calculations on the Mn magnetization dependence of spin Hall effect. The intrinsic spin Hall effect in CuAu-I-type antiferromagnets as determined from \textit{ab initio} calculations has been shown to explain the measured spin Hall angles with satisfactory  quantitative agreement \cite{wz_prl}.  This motivates us to interpret the differences in the measured spin Hall angles between \textit{a}- and \textit{c}-axis grown antiferromagnets in terms of the anisotropy of the intrinsic spin Hall effect. While the \textit{a}- and \textit{c}-axis grown antiferromagnets are well textured in the out-of-plane direction as seen in the x-ray diffraction, we assume them to be only weakly textured in the in-plane direction \textcolor{black}{due to the existence of the Cu dusting layer and the polycrystalline nature of Py that deteriorate any possible in-plane epitaxy of the samples. This assumption is further corroborated} by the finding that neither the spin-torque ferromagnetic resonance lineshape nor the resistivity depend on whether the devices are made parallel to the edge MgO[100] or to the diagonal MgO[110] directions of the substrate. We therefore carried out \textit{ab initio} calculations of the intrinsic spin Hall conductivity where we performed  a polycrystalline averaging over the in-plane orientation of the crystals (see Ref. [\onlinecite{wz_prl}] for computational details).  

\begin{table}[b]
\begin{tabular*}{2.9in}[t]{llllllllll}
\hline \\
&growth
&\textcolor{black}{$d_0 (\AA)$}
&$M_\mathrm{stag}$
&$\sigma^{\rm SHE}$ 
&$\sigma^{\rm SHE}_{\rm av}$ 
&$\sigma^{\rm SHE}_{\rm exp}$ 
&\textcolor{black}{$\rho$ ($\mu \Omega$ cm)}
\\ \hline
\multirow{4}{*}{PtMn}  &$c$-axis &\multirow{2}{*}{\textcolor{black}{3.67}} &$c$ & 219.9 &\multirow{2}{*}{94.2} &\multirow{2}{*}{144} &\multirow{2}{*}{\textcolor{black}{180.3}}\\
                      &$c$-axis & &$a/b$ & 31.4  & & \\ 
                      &$a$-axis &\multirow{2}{*}{\textcolor{black}{4.00}}  &$c$ & 182 &\multirow{2}{*}{141} &\multirow{2}{*}{263} &\multirow{2}{*}{\textcolor{black}{163.4}}\\
                      &$a$-axis & &$a/b$ & 120  & & \\ \hline
\multirow{4}{*}{IrMn} &$c$-axis &\multirow{2}{*}{\textcolor{black}{3.64}} &$c$ & 59.7 &\multirow{2}{*}{93.5} &\multirow{2}{*}{77.9} &\multirow{2}{*}{\textcolor{black}{320.8}} \\
                      &$c$-axis & &$a/b$ &110.4  & & \\
                      &$a$-axis &\multirow{2}{*}{\textcolor{black}{3.86}} &$c$ & 207 &\multirow{2}{*}{16} &\multirow{2}{*}{53.2} &\multirow{2}{*}{\textcolor{black}{216.3}}\\
                      &$a$-axis & &$a/b$ &-80  & & \\ \hline
\multirow{4}{*}{PdMn} &$c$-axis &\multirow{2}{*}{\textcolor{black}{3.58}} &$c$ & 17.0 &\multirow{2}{*}{7.0} &\multirow{2}{*}{59.2} &\multirow{2}{*}{\textcolor{black}{270.5}}\\
                      &$c$-axis & &$a/b$ &  2.0  & & \\
                      &$a$-axis &\multirow{2}{*}{\textcolor{black}{4.07}} &$c$ & 44 &\multirow{2}{*}{2.7} &\multirow{2}{*}{99.2} &\multirow{2}{*}{\textcolor{black}{196.6}}\\
                      &$a$-axis & &$a/b$ &-18  & & \\ \hline
\end{tabular*}
\caption{
Calculated spin Hall conductivities $\sigma^{\rm SHE}$ [units in $\frac{\hbar}{e} \frac{S}{\rm cm}$] and comparison to experiment.
}
\label{table_theory_she}
\end{table}

In Table III we list \textcolor{black}{the growth, lattice constant, $d_0$ (unit in $\AA$), direction of staggered magnetization, electrical resistivity, $\rho$ (unit in $\mu \Omega$ cm),} and the resulting intrinsic spin Hall conductivities. Besides the distinction between \textit{a}- and \textit{c}-axis growth the calculated spin Hall conductivities also depend on whether the staggered magnetization is along the \textit{c}-axis or along the \textit{a}- or \textit{b}-axes. The orientation of the staggered magnetization in the thin antiferromagnetic layers is unknown and we assume it to be random along the main crystallographic axes. Therefore, we list also the averages over magnetization directions, defined by:
\bege
\sigma^{\rm SHE}_{\rm av}=
[
2\sigma^{\rm SHE}(a/b-{\rm axis})
+
\sigma^{\rm SHE}(c-{\rm axis})
]/3.
\ee

In the case of PtMn the calculated $\sigma^{\rm SHE}_{\rm av}$ is larger for \textit{a}-axis growth than for \textit{c}-axis growth in agreement with experiment. Good agreement also holds for IrMn, where both theory and experiment find the spin Hall conductivity to be larger for \textit{c}- than \textit{a}-axis growth, opposite to the trend for PtMn.  In the case of PdMn the polycrystalline averaged intrinsic spin Hall conductivities are considerably smaller than experiment, which was also observed in our previous spin pumping experiments \cite{wz_prl}. Further investigations are needed to address the large discrepency for PdMn between experiment and theory.

To further elaborate such anisotropic effect, we choose again PtMn for a more detailed study due to its largest spin Hall angle among all antiferromagnets herein. Figure 6(a) and (b) compare the lineshapes of spin-torque ferromagnetic resonance signals for \textit{a}- and \textit{c}-axis samples. More symmetric over antisymmetric Lorentzian lineshapes can be observed for \textit{a}-axis samples with and without the Cu dusting layer, confirming again minimal interface-induced spin-orbit effects. The individual-$V_\mathrm{s}$ analysis yields 0.048$\pm$0.006 for \textit{c}-axis and 0.089$\pm$0.006 for \textit{a}-axis PtMn, which are similar to values obtained from the ratio analysis. Thus for PtMn we conclude that the magnitudes of spin-Hall effect follow the relationship of \textit{a}-axis $>$ polycrystalline $>$ \textit{c}-axis samples.  

\begin{figure}[t]
\includegraphics[width=1\columnwidth]{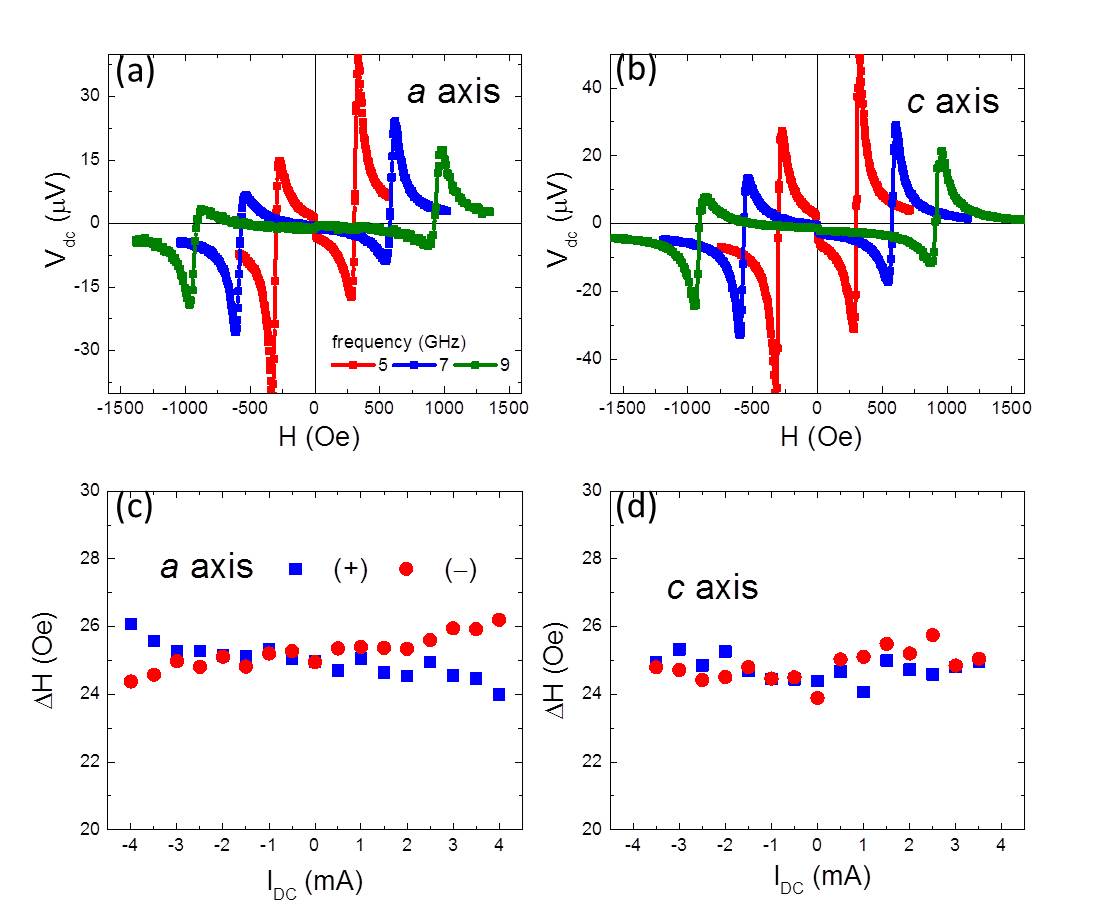}
\caption{\label{fig6} (Color online) Measured spin-torque ferromagnetic resonance signals from (a) \textit{a}- and (b) \textit{c}-axis PtMn(10)/Cu(1)/Py(5), and (c,d) the corresponding resonance linewidth versus dc-bias current at 5 GHz.}
\end{figure}

Figure 6(c) and (d) compare the dc-tuned linewidth for the two different axes. A clear modulation effect can only be observed for \textit{a}- and not for \textit{c}-axis textured samples, confirming much smaller spin Hall effect for the latter. The large spin Hall effect of PtMn originates from the large atomic spin-orbit coupling of Pt, acting as an effective field bending electron trajectory along opposite directions for up and down spins. On the other hand, the staggered magnetization from Mn atoms also indirectly affects their intrinsic spin-orbit coupling via orbital hybridization \cite{kren_pr1968}. The exact mechanism dictating the spin Hall effect of these alloys may require further experimental and theoretical elaborations, but the observed orientation-dependent effects offer a possible route for tunable spin Hall effects in magnetically ordered materials.

\section{\label{sec:level1}Summary}

In summary, we demonstrate spin torque effects of CuAu-I-type antiferromagnets by using spin torque ferromagnetic resonance of Py in combination with a dc-tuned technique. The observed non-local spin torques are attributed to spin currents generated by the spin Hall effect. By using epitaxial samples, we also show the anisotropic spin torque effects upon changing of the growth orientations, which are corroborated by \textit{ab initio} calculations. Our results highlight the important roles of both the heavy elements and the staggered magnetization to the intrinsic spin Hall effects of these alloys.

\section{\label{sec:level1}Acknowledgements}

The experimental work at Argonne was supported by the U.S. Department of Energy, Office of Science, Materials Science and Engineering Division. Lithography was carried out at the Center for Nanoscale Materials, an Office of Science user facility, which is supported by DOE, Office of Science, Basic Energy Science under Contract No. DE-AC02-06CH11357. We gratefully acknowledge computing time on the supercomputers \mbox{JUQUEEN} and \mbox{JUROPA} at J\"ulich Supercomputing Center and the theoretical work at J\"ulich was supported by funding under the HGF-YIG programme VH-NG-513 and SPP 1538 of DFG.

\end{document}